\documentclass[10pt,journal,twocolumn, times,letter]{IEEEtran}
\usepackage{epsfig,times,graphics}
\usepackage{amsmath,amsfonts,amssymb}
\usepackage{cite,graphicx,subfigure,url}
\usepackage{setspace}

\def\done{\hspace*{\fill} \rule{1.8mm}{2.5mm} }

\begin{document}
\title{Understanding the Paradoxical Effects of Power Control on the Capacity of Wireless Networks}

\author{Yue Wang$^{*}$ \hspace{0.3in}\footnote{* Department of Computer Science \& Engineering, The Chinese University of Hong Kong (email: \{ywang, cslui\}@cse.cuhk.edu.hk)}
        John C.S. Lui$^{*}$ \hspace{0.3in} Dah-Ming Chiu$^{+}$\\
        $^{*}$Department of Computer Science \& Engineering\\
        $^{+}$Department of Information Engineering\\
        The Chinese University of Hong Kong \\
        Email: \{{\tt ywang,cslui}\}{\tt@cse.cuhk.edu.hk},
               {\tt dmchiu@ie.cuhk.edu.hk}
}

\maketitle
\begin{abstract}
Recent works show conflicting results: network capacity may increase
or decrease with higher transmission power under different
scenarios. In this work, we want to understand this paradox.
Specifically, we address the following questions: (1)Theoretically,
should we increase or decrease transmission power to maximize
network capacity? (2) Theoretically, how much network capacity gain
can we achieve by power control? (3) Under realistic situations, how
do power control, link scheduling and routing interact with each
other? Under which scenarios can we expect a large capacity gain by
using higher transmission power? To answer these questions, firstly,
we prove that the optimal network capacity is a non-decreasing
function of transmission power. Secondly, we prove that the optimal
network capacity can be increased unlimitedly by higher transmission
power in some network configurations. However, when nodes are
distributed uniformly, the gain of optimal network capacity by
higher transmission power is upper-bounded by a positive constant.
Thirdly, we discuss why network capacity in practice may increase or
decrease with higher transmission power under different scenarios
using carrier sensing and the minimum hop-count routing. Extensive
simulations are carried out to verify our analysis.
\end{abstract}


\section{{\bf Introduction}}\label{sec:intro}
Wireless networks have been actively developed for providing
ubiquitous network access in the past decades. Recently, wireless
mesh networks (WMNs) are considered as a key solution to extend the
coverage of the Internet, especially in areas where wired networks
are expensive to deploy, e.g., in rural areas. Therefore, improving
network capacity is one of the most important issues in the research
of wireless networks. Roughly speaking, network capacity is the
total end-to-end throughputs, which we will carefully define in
Section~\ref{sec:model}. Various techniques ranging from physical
layer to network layer have been proposed for this purpose, such as
MIMO\cite{mimo}, multi-channel multi-radio\cite{mchan},
high-throughput routing \cite{etx,wcett,exor,mtm}, etc.
One way to increase network capacity is by leveraging transmission
power. This is effective especially in WMNs where stationary mesh
routers usually have sufficient power supply, for example, they can
share power supply with street-lamps as cited in\cite{nortel}.

In this paper, we study the impact of power control on the capacity
of wireless networks. In particular, we consider wireless networks
where nodes are stationary and are connected in ad-hoc manner. Under
this network setting, power control can significantly affect network
capacity via the interactions with the link scheduling and the
routing algorithms.

First, many link scheduling algorithms in wireless networks nowadays
implement carrier sensing to avoid transmission collisions due to
interferences\footnote{We do not consider CDMA at the moment, which
applies some other techniques for interference cancellation.}. That
is, transmitters sense channel states before transmissions and they
can transmit only when the sensed noise strength is below carrier
sensing threshold. Power control has a tight relation with carrier
sensing. When transmission power increases, the sensed noise
strength, mainly due to interference, is more likely beyond carrier
sensing threshold, which may reduce spatial reuse, i.e., the number
of simultaneous transmissions. Since network capacity decreases with
lower spatial reuse, higher transmission power may decrease network
capacity.
Second, power control has a tight relation with routing. On the one
hand, higher transmission power may reduce the number of hops or
transmissions that a source needs to reach its destination for a
longer transmission range. Since network capacity increases with
fewer number of transmissions for an application-layer packet,
higher transmission power may increase network capacity. On the
other hand, because longer transmission range reduces spatial reuse
(see Section~\ref{sec:model}), higher transmission power can
decrease network capacity. Considering perfect link scheduling,
authors in \cite{Kumar_power} argued that network capacity decreases
with higher transmission power under the minimum hop-count routing.
However, some recent works showed that network capacity actually
increases with higher transmission power in some scenarios
\cite{load_pc}\cite{high_power}.

In this paper, we systematically characterize the impact of power
control on network capacity and provide a deep understanding on the
interesting \emph{paradox}: why network capacity may increase or
decrease with higher transmission power in different scenarios?
Specifically, we address the following questions:
\begin{enumerate}
\item Theoretically, should we increase or decrease transmission power to maximize network capacity?
\item  Theoretically, how much
network capacity gain can we achieve by power control?
\item  Under realistic situations,
how do power control, link scheduling and routing interact with each
other? Under which scenarios can we expect a large capacity gain
using higher transmission power?
\end{enumerate}
The contributions of this work are as follows:
\begin{itemize}
\item We prove that the \emph{optimal} network capacity is a \emph{non-decreasing} function of transmission power
when the network is using the optimal link scheduling and routing.

\item We prove that under some specific configurations, the optimal
network capacity can be increased unlimitedly by higher transmission
power. However, when nodes are distributed uniformly over a space,
the gain of the optimal network capacity by higher transmission
power is upper-bounded by some positive constant. To the best of our
knowledge, we are the first to prove this property.

\item We provide a qualitative analysis on the interactions of
power control, carrier sensing and the minimum hop-count routing.
The later two are the key features commonly used in the link
scheduling and routing algorithms nowadays. Through this analysis,
we can explain the paradoxical effects of power control on
increasing network capacity. The essential reason is that carrier
sensing and the minimum hop-count routing are not optimal. We also
provide a taxonomy of different scenarios where network capacity
(may) increase or decrease with higher transmission power.

\item Besides the theoretical contributions, our work offers some
important implications to network designers. First, one can redesign
the link scheduling and routing algorithms so as to increase network
capacity under high transmission power. Second, we observe from
simulation that high transmission power can significantly increase
network capacity in the networks whose diameters are within a few
hops, which can find applications in small WMNs.
\end{itemize}

The rest of the paper is organized as follows. In
Section~\ref{sec:model}, we present a model of wireless networks and
define performance measures. In Section~\ref{sec:theorems}, we prove
the theoretical network capacity gain of power control. In
Section~\ref{sec:real}, we discuss why network capacity in practice
may increase or decrease with higher transmission power, considering
the interactions of power control, carrier sensing and the minimum
hop-count routing. In Section~\ref{sec:simulation}, we study how
network capacity varies with transmission power in different
scenarios via simulation. In Section~\ref{sec:related}, we present
related works. In Section~\ref{sec:conclusion} we conclude our
paper.

\section{{\bf System Model}}\label{sec:model}
In this section, we first present a physical model commonly used in
the research of wireless networks \cite{capacity}. Then we define
performance measures and some notations used throughout this paper.

In this paper, we consider a \emph{static} network of $n$ nodes
which are located on a 2D plane. Nodes are connected in ad-hoc
manner. We use $(A, B)$ to denote a link transmitting from node $A$
to node $B$, and use $|A-B|$ to denote the Euclidean distance
between $A$ and $B$.
We make the following assumptions for the wireless physical model:
\emph{1) Common transmission power.} All nodes use the same
transmission power. This assumption simplifies our discussions.
Actually, the authors of the COMPOW (COMmon POWer) protocol showed
that per-node (or per-link) power control can only improve network
capacity marginally than \emph{common power}
control\cite{Kumar_power}. \emph{2) Single ideal channel.} All nodes
transmit on an \emph{ideal} channel without channel fading. This
assumption simplifies our analysis so that we can focus on
understanding this paradox.  In practice, there are some physical
technologies such as MIMO which can greatly mitigate channel fading
by using smart antennas\cite{mimo}. \emph{3) Single transmission
rate.} All nodes transmit at the same date rate of $W$ bps. \emph{4)
Correct packet reception based on signal-to-noise (SNR) threshold.}

Let $P_t$ be the transmission power. For a link $e$, the received
signal strength $P_r$ at $e$'s receiver is
\begin{eqnarray}
P_r = \frac{c_p P_t}{d^\alpha}, \label{eq:rss1}
\end{eqnarray}
where $c_p$ is a constant determined by some physical parameters,
e.g. antenna height, $\alpha$ is the path loss exponent, varying
from $2$ to $6$ depending on the environment\cite{wireless}, and $d$
is the distance from $e$'s transmitter to its receiver (we call it
the \emph{length} of link $e$). We assume all $c_p$'s are equal.
Thus, by letting $P_t$ denote $c_p P_t$, we can simplify
Eq.~(\ref{eq:rss1}) as
\begin{eqnarray}
P_r = \frac{P_t}{d^\alpha}. \label{eq:rss}
\end{eqnarray}

For link $e$, its signal-to-noise (SNR) is defined at its receiver
side, which is
\begin{eqnarray}
SNR = \frac{P_r}{\sum_{i \neq e}{I_i} + N_0}, \label{eq:snr}
\end{eqnarray}
where $P_r$ is the signal strength at $e$'s receiver, $I_i$ is the
interference strength from some other transmitting link $i$ to $e$,
and $N_0$ is the white noise. $I_i$ is also calculated by
Eq.(\ref{eq:rss}) except that $d$ here is the distance from $i$'s
transmitter to $e$'s receiver. The accumulative interference
strength and $N_0$ are treated as \emph{noise} by $e$'s receiver.
Note that $N_0$ is usually small comparing with interference
strength so that we can ignore it.

To successfully receive a packet, the following two conditions
should both be satisfied:
\begin{eqnarray}
P_r \geq H_r, \label{eq:recv_thresh}
\end{eqnarray}
    and
\begin{eqnarray}
SNR \geq \beta, \label{eq:recv_snr}
\end{eqnarray}
where $H_r$ is the receiving power threshold and $\beta$ is the SNR
threshold for decoding packets correctly.

From the above equations, one can derive $r$, the maximum distance
between a transmitter and a receiver for successful packet
receptions (the maximum is achieved when interference is zero),
\begin{eqnarray}
r = \min  \left\{ \left(\frac{P_t}{N_0 \beta}\right)^{1/\alpha},
     \left(\frac{P_t}{H_r}\right)^{1/\alpha} \right\}.
\label{eq:tx_range}
\end{eqnarray}
We refer to $r$ as \emph{transmission range}. Two nodes can form a
link when they are within a distance of $r$.

The \emph{interference range} $r_I$ of a link $e$ is defined as the
minimum distance between an interfering transmitter and $e$'s
receiver so that $e$'s transmissions are not corrupted. Let $d$ be
the length of $e$. From Eq.~(\ref{eq:rss})-(\ref{eq:snr}), and
ignoring $N_0$, we have
\begin{eqnarray}
\frac{P_t/d^\alpha}{P_t/r_I^\alpha} = \beta,\nonumber
\end{eqnarray}
which yields
\begin{eqnarray}
r_I = \beta ^{1/\alpha} \cdot d \label{eq:inf_range}
\end{eqnarray}

We observe that $r_I$ is a constant times of $d$ and is independent
of transmission power. Another observation is that the \emph{silence
area} for successful transmissions of a link is proportional to the
link length. This suggests that spatial reuse, i.e. the number of
simultaneous transmissions, will decrease with the lengths of links.

Next, we define network capacity according to
\cite{mchan_cap}\footnote{We adopt this definition of network
capacity because it isolates the capacity definition from fairness
concerns}, which is from the perspective of end-users. We consider a
network $G$ and a set of flows $F$. Each flow is associated with a
rate. The rate of a flow is the average end-to-end throughput of the
flow. We use a vector to denote the rates of all flows, named
\emph{flow rate vector}. \emph{Capacity region} defines all flow
rate vectors that can be \emph{supported} by the network.

We define \emph{traffic pattern} as the ratio of the rates of all
flows, which can be represented in the vector form: $(v_1, v_2,...,
v_{|F|})$, where $v_1^2 + v_2^2+...+v_{|F|}^2 = 1$. Given the
traffic pattern, we can obtain a corresponding flow rate vector
$a\cdot(v_1, v_2,..., v_{|F|})$ by a scaling factor $a$. The
\emph{network capacity} under the traffic pattern of $(v_1, v_2,...,
v_{|F|})$ is defined as
\begin{eqnarray}
\max_{a>0} \left\{a \cdot \!\!\sum_{i=1...|F|}{v_i}
\right\},\label{eq:cap}
\end{eqnarray}
, which is the maximum total rates of flows supported by the
network.

\begin{figure}[htb]
\centering
    \includegraphics[width=0.10\textwidth]{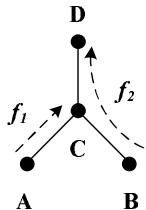}
\caption{Illustration of the definition of network capacity}
\label{fig:capa_eg}
\end{figure}

We illustrate the above definitions by an example. There are four
nodes ($A$, $B$, $C$ and $D$) and two flows ($f_1$ from $A$ to $C$
and $f_2$ from $B$ to $D$) in the network of Fig.~\ref{fig:capa_eg}.
So there are three links ($(A,C)$, $(B,C)$ and $(C,D)$) contending
the channel. Let $\lambda_1$ and $\lambda_2$ be the rates of the two
flows, respectively. We can easily calculate the capacity region of
$(\lambda_1, \lambda_2)$ by the constraint $\lambda_1 + 2\lambda_2
\leq W$. Suppose the traffic pattern is $(\frac{1}{\sqrt{2}},
\frac{1}{\sqrt{2}})$, then the network capacity is $\frac{2}{3}W$
when $\lambda_1 = \lambda_2 = \frac{1}{3}W$.

Equivalently, we can calculate network capacity as follows. Given
the traffic pattern $(v_1, v_2,..., v_{|F|})$, we generate the
corresponding traffic workload vector $b\cdot(v_1, v_2,...,
v_{|F|})$ by a large scaling factor $b$ ($b\cdot v_i$ is the traffic
workload assigned to the $i^{th}$ flow). Suppose that the network
delivers all traffic workloads in time $T$, then the network
capacity is
\begin{eqnarray}
\frac{b \cdot \!\!\sum_{i=1...|F|}{v_i}}{T}. \label{eq:cap1}
\end{eqnarray}

Finally, we define the network capacity gain of power control. Given
the wireless network and the traffic pattern, let $C_P(R, S)$ be the
network capacity when $P_t = P$ under the routing algorithm $R$ and
the link scheduling algorithm $S$. $R$ defines the routes of each
flow, and $S$ defines whether a link can transmit at any time $t$.
We use $C_P^*(R^*, S^*)$ or $C^*_P$ to denote the optimal network
capacity when $P_t = P$ under the optimal routing algorithm $R^*$
and the optimal link scheduling algorithm $S^*$.

Let $P$ and $KP$ ($K>1$) be the minimal and the maximal transmission
power, respectively. Note that $P$ should guarantee \emph{network
connectivity}; Otherwise, network capacity is meaningless since some
flows may not be able to find routes to reach their destinations. We
define \emph{network capacity gain of power control} ($G_K(R, S)$)
by using the routing algorithm $R$ and the link scheduling algorithm
$S$ as
\begin{eqnarray}
G_K(R, S) = \frac{C_{KP}(R, S)}{C_P(R, S)}.
\end{eqnarray}
Furthermore, we define the \emph{theoretical network capacity gain
of power control} ($G^*_K$), i.e.,
\begin{eqnarray}
G^*_K = \frac{C^*_{KP}}{C^*_P}.
\end{eqnarray}
Unless we state otherwise, we will use $K$ to denote the ratio of
the maximal transmission power to the minimal transmission power in
this paper.

\section{{\bf Theoretical network capacity gain of power control}}\label{sec:theorems}

In this section, we derive the theoretical capacity gain of power
control based on the information-theoretic perspective. In order to
derive the optimal network capacity, we assume that nodes transmit
in a synchronous time-slotted mode and each transmission occupies
one time slot. From now on we will use the phrase \emph{"with high
probability"} abbreviated as \emph{"whp"} to stand for \emph{"with
probability approaching $1$ as $n \rightarrow \infty"$} where $n$ is
the number of nodes in the network.

The following theorem states the relationship between the optimal
network capacity and transmission power.

\newtheorem{theorem}{\hspace{-0.15in}{\bf Theorem}}
\newtheorem{lemma}{\hspace{-0.15in}{\bf Lemma}}
\ \\
\begin{theorem}\label{theo:compower}
\emph{Given the network topology and the traffic pattern, the
optimal network capacity is a non-decreasing function of the common
transmission power. Therefore, $G^*_K \geq 1$.}
\end{theorem}

\noindent \textbf{\emph{Proof:}} Let $S^*_P(t)$ denote the set of
transmitting links at time slot $t$ when $P_t = P$. For any link $e
\in S^*_P(t)$, its SNR satisfies
\begin{eqnarray}
\frac{P_r}{\sum_{i\in S_P(t), i\neq e}{I_i}+N_0} \geq \beta,
\end{eqnarray}
where $P_r$ is the signal strength of $e$ and $I_i$ is the
interference strength from some other transmitting link $i$ to $e$.
Now we set $P_t = KP$ ($K > 1$) and use the \emph{same} routes and
the \emph{same} link scheduling sequence as $P_t=P$. We can see that
at time slot $t$, $e$'s SNR is
\begin{eqnarray}
\frac{KP_r}{\sum_{i\in S^*_P(t), i\neq e}{KI_i+N_0}} >
\frac{P_r}{\sum_{i\in S_P(t), i\neq e}{I_i}+N_0} \geq \beta,
\end{eqnarray}
where we use the fact that $P_r$ and $I_i$ are proportional to
$P_t$. So $S^*_P(t)$ can be scheduled at $t$ when $P_t = KP$ for any
$t$. Since $R^*$ and $S^*$ are optimal routing and link scheduling,
we have $C^*_{KP} \geq C^*_P$ by optimality. \done

\noindent \textbf{\emph{Remarks:}} The theorem seems counter
intuitive but is easy to understand. Basically, given a set of
simultaneous links, SNR does not decrease with higher transmission
power because both signal strength and interference strength
increase at the same ratio. Network capacity can be further improved
if we can find \emph{better} routes under higher transmission power.
Therefore, theoretically, it is desirable to use higher transmission
power to increase network capacity.
\ \\

An interesting question is how much network capacity gain we can
achieve by using higher transmission power. To answer this question,
let us analyze it based on the information-theoretic perspective
\cite{capacity}. Without loss of generality, we scale space and
suppose that $n$ nodes are located in a disc of unit area.
\ \\
\begin{theorem}\label{theo:gain_general}
\emph{In general, $G^*_K$ can be unbounded when $n \rightarrow
\infty$}.
\end{theorem}

\begin{figure}[htb]
\centering
    \includegraphics[width=0.5\textwidth]{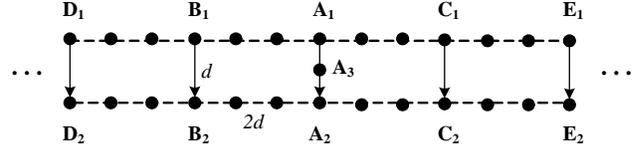}
\caption{A network having unbounded $G^*_K$} \label{fig:unbounded}
\end{figure}

\noindent \textbf{\emph{Proof:}} We prove it by constructing a
specific network. There are $2m+1$ vertical links each with a length
of $d$. The horizontal distance between any two adjacent vertical
links is $2d$. Fig.~\ref{fig:unbounded} illustrates five vertical
links where ($A_1, A_2$) is the middle link of the network. $A_3$
evenly separates the line between $A_1$ and $A_2$. Also, there are
two nodes evenly separating the line between any two horizontally
neighboring nodes. So there are totally $n = 12m+3$ nodes in the
network. There is a flow along each vertical link from the top node
to the bottom node. Let $\alpha = 4$ and $\beta = 10$ in the
physical model.

The maximal transmission power $KP$ is set large enough that the
transmission range $r$ is much larger than $d$ and $N_0$ can be
neglected. Thus, the $2m+1$ vertical links can transmit
simultaneously for any $m$. To see this, we can check the SNR of the
middle link ($A_1, A_2$) which suffers the most interference, i.e.,
\begin{eqnarray}
SNR_{(A_1, A_2)} &\geq&
\frac{\frac{KP}{d^4}}{2\cdot\sum_{i=1}^{m}\frac{KP}{(\sqrt{d^2+(2id)^2})^4}}. \nonumber\\
\end{eqnarray}
$SNR_{(A_1, A_2)} \approx 11 > \beta$ when $m \rightarrow \infty$.
Therefore, $C^*_{KP}$ is $(2m+1)W$ or $(\frac{1}{6}n+\frac{1}{2})W$.
The minimal transmission power $P$ is set so that $d > r >
\frac{2}{3}d$. Thus all flows have to go through $A_1$, $A_3$ and
$A_2$ to reach their destinations. For example, the route from $E_1$
to $E_2$ is through $C_1$, $A_1$, $A_3$ $A_2$ and $C_2$. So $C^*_P$
is at most $\frac{1}{2}W$ since $(A_1, A_3)$ and $(A_3, A_2)$ are
the bottleneck links for all flows. Therefore, $G^*_K$ is at least
$(\frac{1}{3}n+1)$, which is unbounded when $n \rightarrow \infty$.
\done

\noindent \textbf{\emph{Remarks:}} The above theorem shows that
network capacity can be increased unlimitedly by using higher
transmission power in some network configurations.

However, nodes placement is approximately random in many real
networks. We will show that $G^*_K$ is upper-bounded by a constant
\emph{whp} for networks with uniform node distribution. Before we
finally prove this result, we have the following lemmas. We first
cite a lemma which was proved in \cite{capacity}.
\ \\
\begin{lemma}\label{lemma:disjoint_links}
\emph{For any two simultaneous links $(A, B)$ and $(C, D)$, we have
$|B-D| \geq \frac{\Delta}{2}(|A-B|+|C-D|)$, where $\Delta =
\beta^{1/\alpha}-1$.}
\end{lemma}

\noindent \textbf{\emph{Remarks:}} From this lemma, if we draw a
disc for each link where the center of the disc is the link's
receiver and the radius is $\frac{\Delta}{2}$ times the link length,
all such discs are disjoint. Note that $\Delta > 0$ because we
usually have $\beta > 1$ in practice.
\ \\

\begin{lemma}\label{lemma:intersect_links}
\emph{Consider a set of simultaneously transmitting links where the
length of any link is at least $d$. Given a region whose diameter is
$2R$, the number of links intersecting the region is upper-bounded
by $\frac{1}{\Delta^4}(4(\Delta+1)\frac{R}{d}+\Delta+2)^2$, where
$\Delta = \beta^{1/\alpha} - 1$.}
\end{lemma}

\noindent \textbf{\emph{Proof:}} See Appendix~A-1.\done
\ \\

We define $r_c$ as the \emph{critical} transmission range for
network connectivity \emph{whp}. From \cite{capacity}, we know that
$r_c = \sqrt{\frac{\log{n}+k_n}{\pi n}}$ for $n$ nodes uniformly
located in a disc of unit area, where $k_n \rightarrow \infty$ as $n
\rightarrow \infty$.
\ \\
\begin{lemma}\label{lemma:divide_links}
\emph{Assume transmission power is sufficiently large so that $r
> 4 r_c$. For a network with uniform node distribution, there exists
a route between any two nodes $A$ and $B$ which satisfies the
following conditions whp: (a) for any relay link on the route, its
length is smaller than or equal to $4 r_c$; (b) the vertical
distance from any relay node to the straight-line segment of $(A,
B)$ is at most $r_c$; (c) the number of hops between any two relay
nodes $a_1$ and $a_2$ is not more than $\frac{|a_1-a_2|}{2 r_c}+1$}.
\end{lemma}

\noindent \textbf{\emph{Proof:}} See Appendix~A-2.\done

\noindent \textbf{\emph{Remarks:}} Intuitively, the lemma shows that
there exists a route which can "approximate" the straight-line
segment of any two nodes \emph{whp} for a network with uniform node
distribution.
\ \\

\begin{theorem}\label{theo:gain_random}
\emph{Assume $\alpha > 2$ and transmission power is sufficiently
large so that $r > 4r_c$. For a network with uniform node
distribution, $G^*_K$ is bounded by a constant $c$ whp, where $c$ is
not depending on $K$ or traffic pattern}.
\end{theorem}

\noindent \textbf{\emph{Proof:}} Let $P$ and $KP$ ($K>1$) be the
minimal and maximal transmission power, respectively. Let
$S^*_{KP}(t)$ be the set of simultaneously transmitting links at
time slot $t$ when $P_t=KP$. To prove this theorem, it is sufficient
to prove that for any $t$ we can schedule the traffic in
$S^*_{KP}(t)$ in at most $c$ time slots when $P_t = P$. By
optimality, we have $G^*_K \leq c$. We will construct such $c$.

To avoid confusion here, we use \emph{"link"} to denote a link when
$P_t = KP$ and use \emph{"sublink"} to denote a link when $P_t = P$.
\emph{Note that} we construct all sublinks from their corresponding
links in this proof according to Lemma~\ref{lemma:divide_links}.
That is, suppose $P$ is sufficiently large so that $r > 4 r_c$, we
can find the \emph{relay sublinks} which satisfy the conditions of
Lemma~\ref{lemma:divide_links} for each link in $S^*_{KP}(t)$
\emph{whp} when $P_t = P$.

First, we will show that such a sublink is interfered by at most
$c_0$ sublinks, where $c_0$ is a constant not depending on $K$ or
traffic pattern. Note that we only consider the links in
$S^*_{KP}(t)$ with a length larger than or equal to $r_c$ here,
since we can schedule the links in $S^*_{KP}(t)$ with a length
smaller than $r_c$ using another time slot.

\begin{figure}[htb]
\centering
    \includegraphics[width=0.45\textwidth]{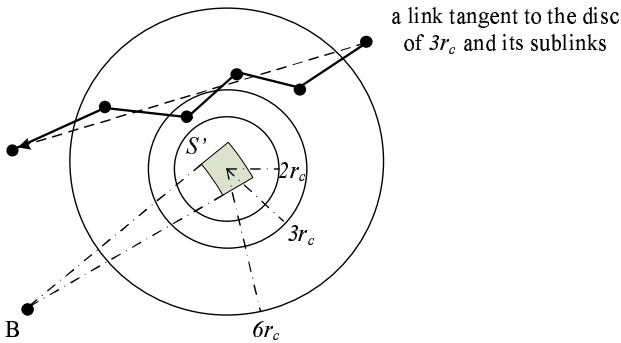}
\caption{Illustration of the worst case for a link whose sublinks
can intersect $S^\prime$} \label{fig:intersect_sector_links}
\end{figure}

We consider some relay sublink $(A, B)$. In the preparatory step, we
count the number of sublinks intersecting the annulus $U(i)$ of all
points lying within a distance between $ir_c$ and $(i+1)r_c$ from
$B$, where $i \geq m$ ($m$ is a constant which we will determine
later). We evenly divide $U(i)$ into $\lceil 2\pi(i+1) \rceil$
sectors, each of which has a central angle of at most
$\frac{1}{i+1}$. Consider such a sector $S$. It is easy to see that
its diameter is not more than $2r_c$. So we can draw a disc of
radius $2 r_c$, named $S^\prime$, to cover $S$. From
Lemma~\ref{lemma:divide_links}, a relay sublink deviates from its
corresponding link by a distance of not more than $r_c$. Therefore,
if a sublink intersects $S^\prime$, the shortest distance between
its corresponding link and $S^\prime$ is at least $r_c$.
Fig.~\ref{fig:intersect_sector_links} illustrates the worst case for
a link (denoted by the directional dashed line) whose sublinks
intersect $S^\prime$, where the link should at least intersect a
disc of radius $3r_c$. Since we consider the links with a length not
less than $r_c$, from Lemma~\ref{lemma:intersect_links}, \emph{the
number of links whose sublinks intersect $S^\prime$} is
upper-bounded by $\frac{1}{\Delta^4}(4(\Delta+1)\frac{3
r_c}{r_c}+\Delta+2)^2 = \frac{1}{\Delta^4}(13\Delta+14)^2$.

A sublink cannot intersect $S^\prime$ if the shortest distance
between its transmitter (or receiver) and $S^\prime$ is larger than
$4r_c$, since its length is not more than $4 r_c$ according to
Lemma~\ref{lemma:divide_links}. Therefore, for any link, \emph{the
number of its corresponding sublinks intersecting $S^\prime$} is
upper-bounded by $\frac{2(2+4)r_c}{2r_c} + 1 = 7$.

From the above results, the number of sublinks intersecting the
annulus $U(i)$ is upper-bounded by $\lceil 2\pi(i+1)\rceil \cdot
\frac{1}{\Delta^4}(13\Delta+14)^2 \cdot 7 < c_1(i+2)$, where $c_1 =
\frac{14\pi}{\Delta^4}(13\Delta+14)^2$. Besides, for a sublink
intersecting $U(i)$, the distance from its transmitter to $B$ is not
less than $(i-4)r_c$. As a result, the total interference to $B$
contributed by the sublinks intersecting $U(i)$ is upper-bounded by
$c_1(i+2) \cdot \frac{P}{((i-4)r_c)^\alpha}$.

Consider the disc $C(B, mr_c)$ of all points lying within a distance
$mr_c$ from $B$. Suppose that \emph{no simultaneous} transmissions
of the sublinks intersecting $C(B, mr_c)$ are allowed, the SNR of
$(A, B)$ is lower-bounded by
\begin{eqnarray}
&&\frac{\frac{P}{(4r_c)^\alpha}}{\sum_{i=m}^\infty c_1(i+2) \cdot
\frac{P}{((i-4)r_c)^\alpha} + N_0}\nonumber\\
&=& \frac{\frac{P}{(4r_c)^\alpha
N_0}}{(\frac{6}{\alpha-1}m^{1-\alpha}+\frac{1}{\alpha-2}m^{2-\alpha})
\cdot\frac{c_1 P}{r_c^\alpha N_0}+1}.\label{eq:theo3_snr}
\end{eqnarray}
We see that the denominator of the last term above approaches $1$
when $m \rightarrow \infty$ for $\alpha > 2$ (In practice, we
usually have $\alpha> 2$ \cite{wireless}. And $\alpha = 2$
corresponds to the free-space path loss model). Suppose $P$ is
sufficiently large so that $r > 4 r_c$, then we have
$\frac{P}{(4r_c)^\alpha N_0}
> \beta$. So there must exist some constant $m$ making Eq.~(\ref{eq:theo3_snr}) larger than or equal to $\beta$.
Clearly, $m$ only depends on $c_1$. Therefore, $(A,B)$ is
\emph{only} interfered by the sublinks intersecting $C(B, mr_c)$. So
the number of sublinks interfering $(A, B)$ is upper-bounded by
\begin{eqnarray}
c_0&=&\frac{1}{\Delta^4}(4(\Delta\!+\!1)\frac{(m+1)r_c}{r_c}\!+\!\Delta\!+\!2)^2
\!\cdot\! (\frac{2(m\!+\!4)r_c}{2r_c}\!+\!1)\nonumber\\
&=&\frac{m+5}{\Delta^4}((4m+5)\Delta+4m+6)^2,
\end{eqnarray}
following the similar arguments above. Note that $c_0$ is not
depending on $K$ or traffic pattern.

Second, we can consider each sublink as a \emph{vertex}. If a
sublink is not interfered by some other sublink, they are assigned
by different \emph{colors}. From the well-known result of vertex
coloring in graph theory, we know that each sublink can be scheduled
at least once in every $c_0 + 1$ slots to finish the traffic of
$S^*_{KP}(t)$.

Finally, consider the links in $S^*_{KP}(t)$ with its length smaller
than $r_c$, we have $c = c_0+2$, where $c$ is not depending on $K$
or traffic pattern. \done

\noindent \textbf{\emph{Remarks:}} First, the assumption of "the
transmission power is sufficiently large" is necessary for $G^*_K$
to be upper-bounded. We illustrate it by an example. Consider there
is one flow transmitting from $A$ to $B$ in a linear topology.
Suppose there is a direct communication between $A$ and $B$ when
$P_t = KP$. So $C^*_{KP}=W$. Suppose there are $m$ hops from $A$ to
$B$ and each hop distance is exactly $r$ when $P_t = P$, where $r$
is the transmission range and $r = (\frac{P}{N_0 \beta})^{1/\alpha}$
(theoretically, we can assume $H_r$ is arbitrarily small).
Obviously, only one hop can transmit successfully at a time to
satisfy the SNR requirement. So $C^*_{P}=\frac{W}{m}$. Therefore
$G^*_K = m$ which is unbounded when $m \rightarrow \infty$. Second,
the assumption of "uniform node distribution" is not necessary for
$G^*_K$ to be upper-bounded. Actually, we can derive the same result
in Theorem~\ref{theo:gain_random} if Lemma~\ref{lemma:divide_links}
holds for some other random node distribution, or more generally, if
the route between any two nodes can "approximate" the straight line
segment of them.
\ \\

In summary, the optimal network capacity is a \emph{non-decreasing}
function of transmission power. Under some specific configurations,
the optimal network capacity can be increased unlimitedly by higher
transmission power. However, when nodes are distributed uniformly
over a space, the gain of optimal network capacity by higher
transmission power is upper-bounded by some positive constant
\emph{whp}.

\section{{\bf Practical Network Capacity Gain of Power Control}}\label{sec:real}
In the previous section we see that network capacity is maximized
under the settings of maximal transmission power, optimal routing
and link scheduling. However, the latter two are NP-hard
problems\cite{inf_impact}\cite{sched_complex}. In this section, we
examine $G_K$ by using carrier sensing and the minimum hop-count
routing, which are the key features commonly used in the link
scheduling and routing algorithms nowadays.

First, we discuss carrier sensing. To avoid collisions during
transmissions, many current solutions require transmitters to sense
channel before transmissions. A transmitter can transmit only when
\begin{eqnarray}
P_{s} \leq H_{s}, \label{eq:cs}
\end{eqnarray}
where $P_s$ is the noise strength sensed at transmitter side and
$H_s$ is \emph{carrier sensing threshold}. Assume the network is
\emph{symmetric}, that is, $P_s$ at transmitter side is equal to
$\sum{I_i} + N_0$ at receiver side (Note that the assumption is
often invalid in practice). By setting $H_s = \frac{P_r}{\beta}$,
one can guarantee that $SNR \geq \beta$ \cite{adapt_cs}. However, it
is difficult in practice for a transmitter to know its $P_r$ at
receiver side. To circumvent this problem, we can conservatively
estimate $P_r$ by $H_r$. So we have
\begin{eqnarray}
H_s = \frac{H_r}{\beta}. \label{eq:cs_thresh}
\end{eqnarray}
$H_s$ in current settings is more or less this value, e.g. Lucent
ORiNOCO wireless card\cite{lucent_card}.

For better illustrations, we introduce \emph{carrier sensing range}
$r_s$, which is defined as the maximum distance that the transmitter
can sense the transmissions of an interfering transmitter. From
Eq.~(\ref{eq:rss}) by letting $P_r = H_s$, we have
\begin{eqnarray}
r_s = \big(\frac{P_t}{H_s}\big)^{1/\alpha}. \label{eq:cs_range}
\end{eqnarray}

Suppose $H_r \geq \beta N_0$, which is usually the case in practice
\cite{802.11a}. From Eq.~(\ref{eq:tx_range}), (\ref{eq:cs_thresh})
and (\ref{eq:cs_range}), we have
\begin{eqnarray}
r_s = \beta^{1/\alpha} \cdot r. \label{eq:cs_range2}
\end{eqnarray}
Comparing with Eq.~(\ref{eq:inf_range}), we see that $r_s$ is equal
to the interference range of the maximum link length.

\begin{figure}[tb]
\centering
    \includegraphics[width=0.45\textwidth]{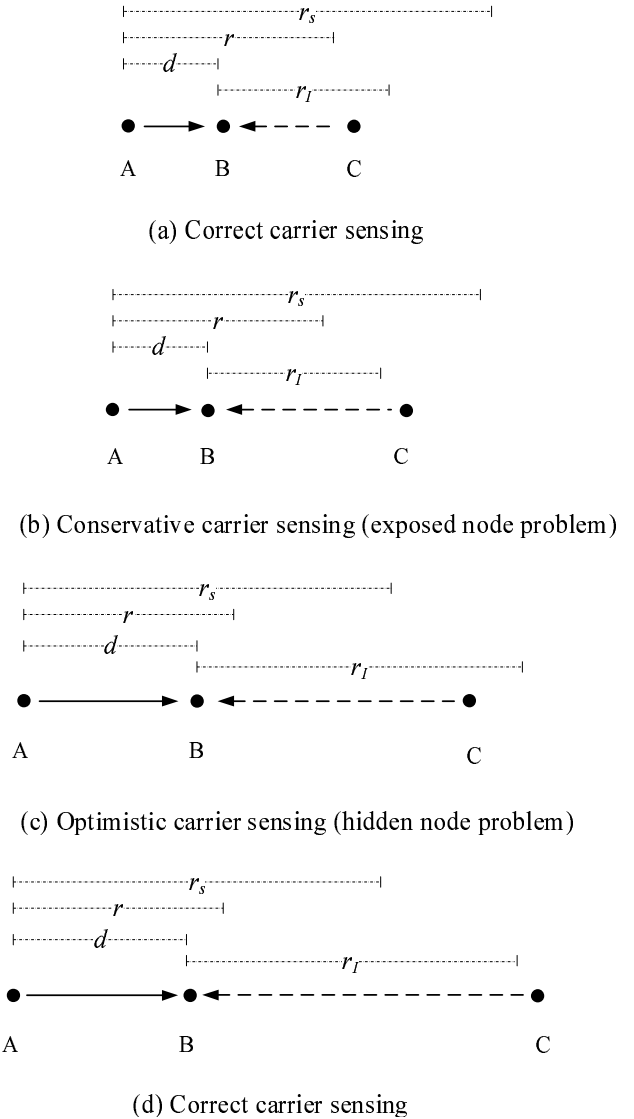}
\caption{Illustration of the relationships of $r$, $r_I$ and $r_s$}
\label{fig:graphic}
\end{figure}

Fig.~\ref{fig:graphic} illustrates the relationships of $r$, $r_I$
and $r_s$ by a network of a transmitter $A$, a receiver $B$ and a
interfering transmitter $C$. Here, we use $d$ to denote $|A-B|$. The
network is not symmetric as $A$ is further from $C$ than $B$ is. In
Fig.~\ref{fig:graphic}(a), $C$ causes packet collisions of $(A,B)$
as it is within $r_I$ of $B$. However, $C$ is also within $r_s$ of
$A$. So $A$ will not transmit and thus avoid collisions when it
senses the transmissions of $C$. In Fig.~\ref{fig:graphic} (b), $C$
is moved outside $r_I$ of $B$ and thus becomes a non-interfering
transmitter to $(A,B)$. So $A$ and $C$ can transmit simultaneously.
However, carrier sensing forbids the simultaneous transmissions as
the $C$ is within $r_s$ of $A$. This case is often referred to as
\emph{exposed terminal (node) problem}. Fig.~\ref{fig:graphic} (c)
and (d) illustrate the scenarios when we increase $d$. By
Eq.~(\ref{eq:inf_range}), $r_I$ also increases and it is not fully
covered by $r_s$ here. In Fig.~\ref{fig:graphic}(c), there will be a
lot of collisions for $(A,B)$ as $C$ is inside $r_I$ of $B$ and
outside $r_s$ of $A$. This case is often referred to as \emph{hidden
terminal (node) problem}. Currently, some MAC protocols (e.g.
802.11) use the backoff mechanism to reduce collisions in this case.
In Fig.~\ref{fig:graphic}(d), $C$ is moved outside $r_I$ of $B$ and
becomes a non-interfering transmitter to $(A,B)$. So $A$ and $C$ can
transmit simultaneously.

Exposed terminal problem is liable to occur when the length of a
link is small, while hidden terminal problem is liable to occur when
the length of a link is large. The radical reason is that carrier
sensing uses fixed $H_s$ and operates at transmitter side, which can
not estimate interference accurately.
%

Therefore, even under the optimal routing, network capacity can
degrade with higher transmission power by using carrier sensing. For
example, consider a network with all one-hop flows, higher
transmission power increases $r_s$, which can reduce spatial reuse
and thus decrease network capacity.

However, the current $H_s$ may not be too conservative under the
minimum hop-count routing, because this kind of routing prefers the
links of longest lengths (approaching $r$), which is close to the
case when we derive Eq.~(\ref{eq:cs_thresh}). Consider a link with a
length $d$, the range that $r_s$ cannot cover $r_I$ is
\begin{eqnarray}
d + r_I - r_s = d + \beta^{1/\alpha} d - \beta^{1/\alpha} r,
\end{eqnarray}
which is approximately $r$ when $d \approx r$. This implies that
there can be more hidden terminals when $r$ becomes larger under the
minimum hop-count routing.

Next, we discuss the minimum hop-count routing. The authors of
\cite{Kumar_power} argued that even under optimal link scheduling
\emph{network capacity by using the minimum hop-count routing} is
proportional to
\begin{eqnarray}
\frac{1}{r}. \label{eq:Kumar_power}
\end{eqnarray}
So $G_K = (\frac{1}{K})^{1/\alpha}$ by Eq.~({\ref{eq:tx_range}).
Their interpretation is as follows. The network capacity consumption
of a flow is proportional to the number of hops the flow traverses,
i.e. $\frac{1}{r}$. Spatial reuse is proportional to
$\frac{1}{r^2}$. Network capacity is proportional to spatial reuse
and \emph{inversely} proportional to the network capacity
consumption per flow, i.e. $\frac{1}{r}$.

We make some comments on Eq.~(\ref{eq:Kumar_power}). First, although
it properly characterizes the \emph{order} of network capacity as a
function of $r$, it has some deviations from practice. For example,
the network diameter (in term of the number of hops) may be so small
that the spatial reuse may not decrease as much as $\frac{1}{r^2}$
due to edge effect\footnote{In here, the edge effect means that the
network diameter is so small that most links are near the periphery
of the network}. As a result, the network capacity may increase with
larger $r$. Fig.~\ref{fig:small_num_nodes} shows an example where
there are five nodes and two flows of equal rate in the network.
When the transmission power is low, both flows need to traverse the
centered node to reach their respective destinations. Since there
are four links contending the channel, the network capacity is
$\frac{1}{4}W \cdot 2 = \frac{1}{2}W$. When we increase the
transmission power so that packets can be transmitted directly from
sources to destinations, there are two links contending the channel,
and network capacity is $\frac{1}{2}W \cdot 2 = W$. Actually, the
spatial reuse here is always one transmitting link per time slot for
any power level due to edge effect. The network capacity increases
with higher transmission power due to a less number of hops per
flow. Second, it \emph{may not} hold for the networks with
\emph{non-uniform link load distribution}. Fig.\ref{fig:non-uniform}
shows an example where there are $k$ flows of equal rate traversing
through the centered node. The link load distribution is non-uniform
here as the centered node is the biggest bottleneck. It is easy to
see that the spatial reuse decreases as $\frac{1}{r^2}$ here.
However, the network capacity does not decrease as $\frac{1}{r}$. To
see this, we consider two specific cases. In the first case of using
the minimal transmission power, each flow is $m$-hop ($m >> 2$). So
there are at least $2k$ links neighboring the centered node,
resulting in the network capacity of at most $\frac{W}{2k} \cdot k =
\frac{1}{2}W$. In the second case of using the maximal transmission
power, each flow is $1$-hop. So there are $k$ links contending the
channel, resulting in the network capacity of $\frac{W}{k} \cdot k =
W$.

\begin{figure}[htb]
\centering
    \includegraphics[width=0.2\textwidth]{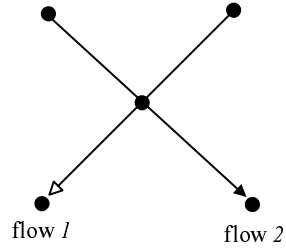}
\caption{An example of a network with a small network diameter}
\label{fig:small_num_nodes}
\end{figure}

\begin{figure}[htb]
\centering
    \includegraphics[width=0.3\textwidth]{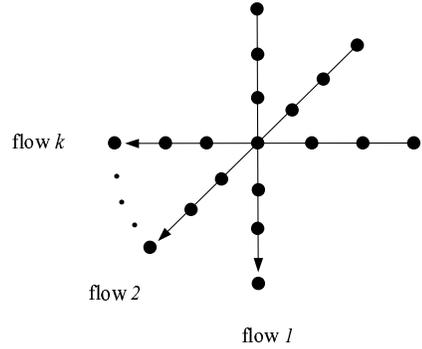}
\caption{An example of a network with non-uniform load distribution}
\label{fig:non-uniform}
\end{figure}

Based on the above observations, one can explain why network
capacity sometimes increases with higher transmission power under
the minimum hop-count routing\cite{load_pc}.

In summary, current carrier sensing and the minimum hop-count
routing do not guarantee $G_K \geq 1$ and may lead to significant
capacity degradation with higher transmission power. However,
network capacity may increase significantly with higher transmission
power in some scenarios, e.g. in networks whose diameter is within a
small number of hops. Therefore, there is a paradox on whether to
use higher transmission power to increase network capacity in
practice.

\section{{\bf Simulation Results}}\label{sec:simulation}
In this section, we examine the impact of power control on network
capacity via simulation. We use carrier sensing and the minimum
hop-count routing as the link scheduling and routing algorithms in
our simulations. Our essential goals are to verify our analysis in
the previous section and to find out under which scenarios we can
expect a large network capacity gain by using high transmission
power.

We use the wireless physical model described in
Section~\ref{sec:model}. We set $\alpha = 4$ for simulating the
two-ray ground path loss model\cite{wireless}. We set $\beta = 10$
and $H_r = -81$dBm\cite{802.11a}. Therefore, $H_s = \frac{1}{10}H_r$
by Eq.~(\ref{eq:cs_thresh}). We ignore $N_0$ which is usually much
smaller than the interference strength. For better illustrations, we
use the transmission range $r$ to represent the transmission power.
We increase the transmission power so that $r=250$m, $500$m, $750$m
and $1000$m. Actually, one can change $r$ proportionally and scale
network topologies at the same time to obtain the similar simulation
results.

We implemented a TDMA simulator for performance evaluation. That is,
nodes transmit in synchronous time-slotted mode and each DATA
transmission and its ACK occupies one time slot. Transmitters sense
the channel one by one at the beginning of each time slot. A
transmitter will transmit a DATA packet when $P_s \leq H_s$ and its
backoff timer expires. The receiver returns an ACK to the
transmitter when it receives the packet successfully. If the
transmitter does not receive an ACK due to packet collision, it will
carry out the exponential backoff. The backoff mechanism is similar
to that of 802.11 except that we backoff the time slot here.

We calculate network capacity according to Eq.~(\ref{eq:cap1}). We
assign a traffic workload to each flow before simulations start and
measure the duration until all flows finish delivering its traffic
workload. In our simulations, each flow has a equal traffic workload
of $500$ equal-sized packets. We generate CBR traffic for each flow
until completing its traffic workload. The CBR rate is set large
enough to saturate the network. Besides, the packet buffer in each
node is set sufficiently large since we do not consider queue
management at the moment.

There are other factors that affect network capacity in practice
such as sophisticated collision resolution mechanisms, TCP
congestion control and queue management. However, by isolating these
factors, we can better understand the key roles of carrier sensing
and the minimum hop-count routing on network capacity.

For simplicity, in the following experiments, we use CS to denote
carrier sensing and use HOP to denote the minimum hop-count routing.
We implemented a centralized link scheduling, named Cen, as a
benchmark, which schedules links one by one in a centralized and
collision-free way and thus ensures maximal spatial reuse. In each
experiment, we take the average of all simulation results for ten
networks.

In the first experiment, we study the interaction of power control
and carrier sensing by considering one-hop flows so as to isolate
the interaction of routing.

\noindent \textbf{Experiment 1 } Network capacity vs Power in a
random network with \emph{one-hop flows}. There are $n=200$ nodes
uniformly placed in a square of $3000$m$\times$$3000$m, which form a
connected network when $r = 250$m. Each node randomly communicates
with one of its nearest neighbors.

Fig.~\ref{fig:nc_1hop} shows the network capacity as a function of
$r$. Obviously, the network capacity by using Cen is almost a
constant in this scenario. However, when we use CS, higher
transmission power causes more exposed terminals and decrease
network capacity, since the carrier sensing threshold is fixed.

\begin{figure}[htb]
\centering
    \includegraphics[width=0.3\textwidth]{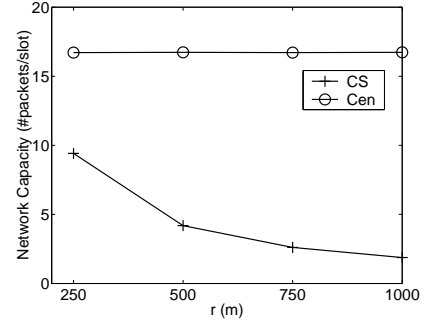}
\caption{Experiment 1: Network capacity as a function of $r$}
\label{fig:nc_1hop}
\end{figure}

In the following experiments, we study the interaction of power
control, carrier sensing and the minimum hop-count routing by
considering multi-hop flows.

\noindent \textbf{Experiment 2 } Network capacity vs Power in a
random network with \emph{multi-hop flows} and \emph{small network
diameter} (in terms of the number of hops). There are $n=20$ nodes
uniformly placed in a square of $1000$m$\times$$1000$m, which form a
connected network when $r = 250$m. Each node randomly communicates
with any other node in the network.

Fig.~\ref{fig:small}(a) shows the network capacity as a function of
$r$. First, in a sharp contrast to Eq.~(\ref{eq:Kumar_power}), the
network capacity by using HOP significantly increases with $r$. The
reason is that the network diameter is so small (4-6 hops) that the
spatial reuse only decreases slightly with larger $r$, as shown in
Fig.\ref{fig:small}(b). Actually, only a few links can transmit
simultaneously in this scenario due to edge effect. HOP minimizes
the number of hops that flows traverse, as shown in
Fig.\ref{fig:small}(c), which is the dominant factor for the
significant increase of network capacity. Second, CS works
reasonably well in this experiment, as compared with Cen (see
Fig.\ref{fig:small}(b)). The reason is that HOP prefers longest
forwarding links for multi-hop flows, which is close to the case
that we derive $H_s$ in Eq.~(\ref{eq:cs_thresh}).

\begin{figure}[htb]
\centering \subfigure[Network capacity as a function of $r$]{
\includegraphics[width=0.3\textwidth]{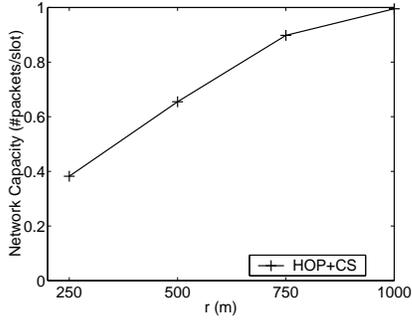}}
\centering \subfigure[Avg. spatial reuse as a function of $r$]{
\includegraphics[width=0.3\textwidth]{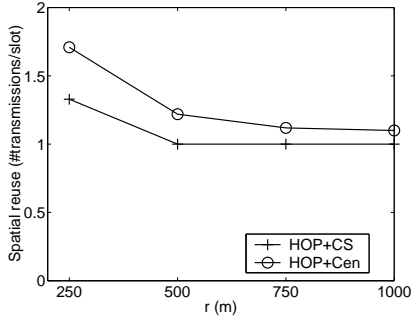}}
\centering \subfigure[Avg. number of hops per flow as a function of
$r$]{
\includegraphics[width=0.3\textwidth]{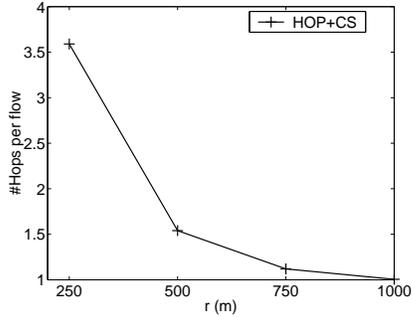}}
\caption{Experiment 2} \label{fig:small}
\end{figure}

\noindent \textbf{Experiment 3 } Network capacity vs Power in a grid
network with \emph{multi-hop flows} and \emph{large network
diameter} (in terms of the number of hops). There are $n=625$ nodes
placed in a $25\times25$ grid. There is a distance of $200$m between
any two horizontally or vertically neighboring nodes. There are $25$
flows from the leftmost nodes to the rightmost nodes horizontally
and $25$ flows from the topmost nodes to the bottommost nodes
vertically. This configuration ensures a large network diameter and
uniform link load distribution.

We observe that the network capacity decreases significantly with
larger $r$, as shown in Fig.~\ref{fig:grid_nc}, because of the
significant decreasing of spatial reuse under the minimum hop-count
routing. We also plot the network capacity by using HOP and Cen,
which confirms our explanation.

\begin{figure}[htb]
\centering
    \includegraphics[width=0.3\textwidth]{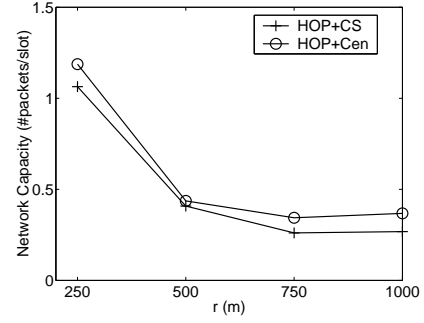}
\caption{Experiment 3: Network capacity as a function of $r$}
\label{fig:grid_nc}
\end{figure}

We also test the random networks with multi-hop flows and a large
network diameter. We observe that the network capacity significantly
decreases with larger $r$ in this scenario when $n$ is sufficiently
large.

In summary, the following conclusions can be made from our analysis
(Section~\ref{sec:real}) and simulations. When we use carrier
sensing and the minimum hop-count routing,
\begin{itemize}
\item In the networks with \emph{one-hop flows}, the network capacity
significantly decreases with higher transmission power due to
exposed terminal problem.
\item In the networks with \emph{multi-hop flows and a small network diameter of a few hops},
the network capacity can increase significantly with higher
transmission power because the edge effect makes spatial reuse only
decrease slightly with larger $r$. This can find applications in
small WMNs. Currently, many WMNs tend to have a small network
diameter (in term of the number of hops), because the end-to-end
throughput of a flow drops significantly with an increasing number
of hops\cite{capacity} \cite{802.11_WMN}.
\item In the networks of \emph{multi-hop flows and a large network diameter}, there are two
subcases. Under \emph{uniform link load distribution}, the network
capacity decreases significantly with higher transmission power as
shown in Eq.~(\ref{eq:Kumar_power}); Under \emph{non-uniform link
load distribution}, it is hard to make a conclusion. The network
capacity \emph{may} increase with higher transmission power as
illustrated by Fig.~\ref{fig:non-uniform}.
\end{itemize}

\section{{\bf Related Work}} \label{sec:related}
In this section, we present related work and highlight our
contributions.

Research on power control can be classified into two classes: energy
oriented and capacity oriented. The first class of works focus on
energy-efficient power control \cite{emac}\cite{pmac}\cite{paro}.
The application is in mobile ad hoc networks (MANETs) or wireless
sensor networks (WSNs), where nodes have limited battery life. Low
transmission power is preferred here to maximize the throughput per
unit of energy consumption, while maximizing overall network
capacity is the secondary consideration. As a result, their
solutions often achieve moderate network capacity. The second class
of works focus on capacity-oriented power control. The application
is in WMNs where mesh routers have sufficient power supply and
maximizing network capacity is the first consideration.

Authors in \cite{Kumar_power} indicated that network capacity
decreases significantly with higher transmission power under the
minimum hop-count routing and they suggested using the lowest
transmission power to maximize network capacity. There are a lot of
works following this suggestion, e.g. \cite{PCDC}\cite{powmac}, and
they observed capacity improvement by using lower transmission
power. However, there is an opposite argument recently. Park et al
showed via simulation that network capacity sometimes increase with
higher transmission power \cite{load_pc}. Behzad et al formulated
the problem of power control as an optimization problem and proved
that network capacity is maximized by properly increasing
transmission power\cite{high_power}.

We also proved that the optimal network capacity is a non-decreasing
function of common transmission power in a simpler way. Furthermore,
we characterized the theoretical network capacity gain of power
control. Besides, we studied the interactions of power control,
carrier sensing and the minimum hop-count routing. As a result, we
explained the above paradox successfully from both theoretical and
practical perspective. Our work provides a deep understanding on the
\emph{structures} of the power control problem and can be seen as an
extension to \cite{Kumar_power}-\cite{high_power}.

Carrier sensing recently attracts attentions in the area of wireless
networks. Many researchers noticed that carrier sensing can
significantly affect spatial reuse and the current carrier sensing
threshold is not optimal in many cases. Xu et al indicated that
RTS/CTS is not sufficient to avoid collisions and larger carrier
sensing range can help to some extend \cite{rts}. Yang et al showed
that the MAC layer overhead has a great impact on choosing carrier
sensing threshold \cite{cs}. Zhai et al considered more factors on
choosing carrier sensing threshold such as different data rates and
one-hop (or multi-hop) flows\cite{mrate_cs}. They showed that
network capacity may suffer a significant degradation if any of
these factors is not considered properly.  Kim et al revealed that
tuning transmission power has the same effect on maximizing spatial
reuse as tuning carrier sensing threshold\cite{sr}.

There are some works on high-throughput routing recently. ETX uses
expected packet transmission times as the routing metric so as to
filter poor channel-quality links in fading channels \cite{etx}.
WCETT extends ETX for multi-channel wireless networks by also
considering contention time and channel diversity \cite{wcett}. MTM
uses packet transmission duration as the routing metric in
discovering high-throughput routes in multi-rate wireless networks
\cite{mtm}. ExOR takes a different approach which forwards packets
opportunistically in fading channels\cite{exor}.

\section{{\bf Conclusion}}\label{sec:conclusion}
This work thoroughly studies the impact of power control on network
capacity from both theoretic and practical perspective. In the first
part, we provided a formal proof that the optimal network capacity
is a non-decreasing function of common transmission power. Then we
characterize the theoretical capacity gain of power control in the
case of the optimal network capacity. We proved that the optimal
network capacity can be increased unlimitedly with higher
transmission power in some network configurations. However, the
increase of network capacity is bounded by a constant with higher
transmission power \emph{whp} for the networks with uniform node
distribution. In the second part, we analyzed why network capacity
increases or decreases with higher transmission power in different
scenarios, by using carrier sensing and the minimum hop-count
routing in practice. We also conduct simulations to study this
problem under different scenarios such as a small network diameter
vs a large network diameter and one-hop flows vs multi-hop flows.
The simulation results verify our analysis. In particular, we
observe that network capacity can be significantly improved with
higher transmission power in the networks with a small network
diameter, which can find applications in small WMNs.



\begin{center}
    {\bf Appendix A-1: The proof of
    Lemma~\ref{lemma:intersect_links}}
\end{center}

\noindent \textbf{\emph{Proof:}} We can draw a disc ($C_R$) of
radius $R$ to cover the given region. We calculate the number of
simultaneous links intersecting $C_R$. Let $l$ be the length of the
longest link. Obviously, we can draw a disc $C_{R+l}$ to cover all
links intersecting $C_R$, where the center of $C_{R+l}$ is that of
$C_R$ and its radius is $R+l$. By Lemma~\ref{lemma:disjoint_links},
each receiver occupies at least an area of $\frac{1}{4}\pi \Delta^2
d^2$. When $l < \frac{2R+d}{\Delta}$, the number of links in
$C_{R+l}$ is upper-bounded by
\begin{eqnarray}
\frac{\pi (R+(1+\frac{\Delta}{2})l)^2}{\frac{1}{4}\pi \Delta^2 d^2}
\leq \frac{1}{\Delta^4}(4(\Delta+1)\frac{R}{d}+\Delta+2)^2.
\end{eqnarray}
The upper bound of the above equation is obtained when
$l=\frac{2R+d}{\Delta}$.

When $l \geq \frac{2R+d}{\Delta}$, from Eq.~(\ref{eq:inf_range}), we
can easily see that the silence area ($A_l$) of the longest link
covers $C_R$, as illustrated in Fig.~\ref{fig:intersect_links}.
Because all the other simultaneous transmitters should be outside
$A_l$, for any other link intersecting $C_R$, its length is at least
the shortest distance from the circle of $A_l$ to the circle of
$C_R$, which is $(1+\Delta)l - l - 2R = \Delta l - 2R$ in the worst
case. By Lemma~\ref{lemma:disjoint_links}, each receiver occupies an
area of at least $\frac{1}{4} \Delta^2 \pi (\Delta l - 2R)^2$. So
the number of links in $C_{R+l}$ is upper-bounded by
\begin{eqnarray}
\frac{\pi (R+(1+\frac{\Delta}{2})l)^2}{\frac{1}{4} \Delta^2 \pi
(\Delta l - 2R)^2} \leq
\frac{1}{\Delta^4}(4(\Delta+1)\frac{R}{d}+\Delta+2)^2.
\end{eqnarray}
The upper bound of the above equation is obtained when
$l=\frac{2R+d}{\Delta}$. Combining the above two cases of $l$, we
proved this lemma.\done

\begin{figure}[htb]
\centering
    \includegraphics[width=0.3\textwidth]{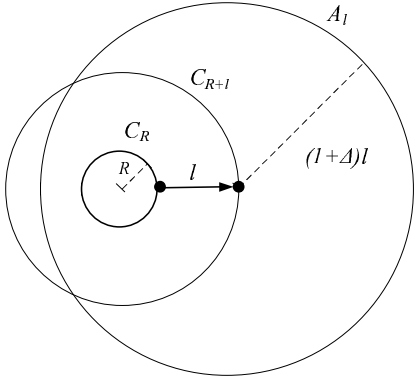}
\caption{Illustration of $C_R$, $C_{R+l}$ and $A_l$}
\label{fig:intersect_links}
\end{figure}

\begin{center}
    {\bf Appendix A-2: The proof of Lemma~\ref{lemma:divide_links}}
\end{center}

\noindent \textbf{\emph{Proof:}} We prove it by constructing such a
route. If $|A-B| \leq 4 r_c$, then $(A,B)$ itself is the desired
route. Otherwise, we divide the straight-line segment of $(A, B)$
into small segments of $2 r_c$ until reaching $B$. Then we draw a
small disc $C_{r_c}(i)$ of radius $r_c$ to cover each small segment,
where $i = 1, 2,...,\lceil \frac{|A-B|}{2r_c}\rceil$.
Fig.~\ref{fig:divide_links} illustrates the case when $\lceil
\frac{|A-B|}{2r_c}\rceil = 3$. For better illustrations, we define
the $x$ axes with its origin at $A$ and its direction from $A$ to
$B$, and define the $y$ axes vertical to $x$. We can see that the
coordinate of the center of $C_{r_c}(i)$ is $((2i-1)r_c, 0)$. The
probability of no node lying in $C_{r_c}(i)$ is $(1-\pi r_c^2)^n$.
Since $|A-B|$ is upper-bounded by the diameter of the disk of unit
area, i.e. $\frac{2}{\sqrt{\pi}}$, the probability that we can
select at least one node in \emph{each} $C_{r_c}(i)$ is
lower-bounded by $(1-(1-\pi r_c^2)^n)^{\frac{2}{\sqrt{\pi} \cdot
2r_c}}$, which approaches $1$ as $n \rightarrow \infty$. Since $r >
4 r_c$, we can connect the selected nodes to form a route from $A$
to $B$. It is easy to see that the route satisfies the conditions of
this lemma.\done

\begin{figure}[htb]
\centering
    \includegraphics[width=0.3\textwidth]{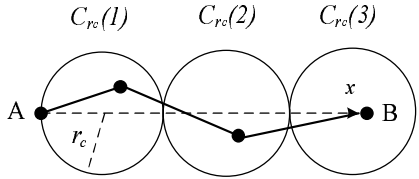}
\caption{Dividing $(A, B)$ into small segments of $2 r_c$}
\label{fig:divide_links}
\end{figure}


\begin{thebibliography}{}
\bibitem{mimo} D. Gesbert, M. Shafi, D. Shiu, P.J. Smith, and A. Naguib, "From
Theory to Practice: An Overview of MIMO Space-Time Coded Wireless
Systems," \emph{IEEE J. Selcted Areas in Comm.}, vol. 21, pp.
281-301, 2003.

\bibitem{mchan} J. Mo, H.-S. W. So, J. Walrand, "Comparison of Multi-channel MAC
Protocols," Procs., \emph{MSWiM}, pp. 209-218, 2005.

\bibitem{nortel} "Nortel: Wireless Mesh Network Solution" , \emph{http://www.nortel.com/solutions/wrlsmesh/}, 2007.

\bibitem{Kumar_power} S. Narayanaswamy, V. Kawadia, R. S. Sreenivas, and P. R. Kumar,
"Power Control in Ad-hoc Networks: Theory, Architecture, Algorithm
and Implementation of the COMPOW Protocol,"  Procs., \emph{European
Wireless Conference}, 2002.

\bibitem{load_pc} S.-J. Park, R. Sivakumar, "Load-Sensitive Transmission Power Control in Wireless Ad-hoc
Networks," Procs., \emph{GLOBECOM}, vol. 1, pp. 42-46, 2002.

\bibitem{high_power} A. Behzad, I. Rubin, "High Transmission Power Increases the Capacity of Ad Hoc Wireless
Networks," \emph{IEEE Trans. on Wireless Comm. }, vol. 5(1), pp.
156-165, 2006.


\bibitem{capacity} P. Gupta, and P. R. Kumar, "The Capacity of Wireless Networks," \emph{IEEE Trans. on Infomation Theory}, vol. 46, no. 2, pp.
388-404, 2000.

\bibitem{mchan_cap} M. Kodialam and T.Nandagopal, "Characterizing the capacity region in
multi-radio multi-channel wireless mesh networks," Procs.,
\emph{MOBICOM}, pp. 73-87, 2005.


\bibitem{wireless} David Tse and Pramod Visvanath, "Fundamentals of Wireless
Communication," \emph{Cambridge University Press}, 2005.

\bibitem{inf_impact} K. Jain, J. Padhye, V. N. Padmanabhan, and L. Qiu, "Impact of Interference
on Multi-hop Wireless Network Performance," Procs., \emph{MOBICOM},
pp. 66-80, 2003.

\bibitem{sched_complex} G. Sharma, R. Mazumdar, and N. Shroff, "On the Complexity of Scheduling in Wireless
Networks," Procs., \emph{MOBICOM}, pp. 227-238, 2006.

\bibitem{adapt_cs} J. Zhu, X. Guo, L. L. Yang, W. S. Conner,
S. Roy, and M. M. Hazra, "Adapting Physical Carrier Sensing to
Maximize Spatial Reuse in 802.11 Mesh Networks," \emph{Wireless
Communications \& Mobile Computing}, vol. 4(8), pp. 933-946, 2004.

\bibitem{lucent_card} "Hardware Specifications of Lucent ORiNOCO Wireless
PC Card," \emph{http://www.orinocowireless.com/.}

\bibitem{802.11a} "IEEE 802.11a. Part 11: Wireless LAN Medium Access Control (MAC) and
Physical Layer (PHY) Specifications: High-speed Physical Layer in
the 5 GHz Band," \emph{Supplement to IEEE 802.11 Standard}, Sep.
1999.

\bibitem{802.11_WMN} J. Bicket, D. Aguayo, S. Biswas, and Robert Morris,
"Architecture and Evaluation of an Unplanned 802.11b Mesh Network,"
Procs., \emph{MOBICOM}, pp. 31-42, 2005.

\bibitem{emac} M. B. Pursley, H. B. Russell, and J. S. Wysocarski,
"Energy-Efficient Transmission and Routing Protocols for Wireless
Multiple-hop Networks and Spread Spectrum Radios," Procs.,
\emph{EUROCOMM}, pp. 1-5, 2000.

\bibitem{pmac} E.-S. Jung and N. H. Vaidya, "A Power Control MAC Protocol for Ad
Hoc Networks," Procs., \emph{MOBICOM}, pp. 36-47, 2002.

\bibitem{paro} J. Gomez, A. T. Campbell, M. Naghshineh, and C. Bisdikian, "PARO:
Supporting Dynamic Power Controlled Routing in Wireless Ad Hoc
Networks," \emph{ACM/Kluwer Journal on Wireless Networks}, vol.
9(5), pp. 443-460, 2003.

\bibitem{PCDC} A. Muqattash, M. Krunz, "Power Controlled Dual Channel (PCDC) Medium Access Protocol for Wireless Ad Hoc
Networks," Procs. \emph{INFOCOM}, vol. 1, pp. 470-480, 2003.

\bibitem{powmac} A. Muqattash and M. Krunz, "A Single-Channel Solution for Transmission Power Control in
Wireless Ad Hoc Networks," Procs., \emph{MOBIHOC}, pp. 210-221,
2004.

\bibitem{rts} K. Xu, M. Gerla, and S. Bae, "How effective is the IEEE 802.11 RTS/CTS
handshake in ad hoc networks," Procs., \emph{GLOBECOM}, 2002.

\bibitem{cs} X. Yang and N. H. Vaidya, "On the Physical Carrier Sensing in
Wireless Ad Hoc Networks," Procs., \emph{INFOCOM}, 2005.

\bibitem{mrate_cs} H. Zhai and Y. Fang, "Physical Carrier Sensing and Spatial Reuse in Multirate and
Multihop Wireless Ad Hoc Networks," Procs., \emph{INFOCOM}, 2006.

\bibitem{sr} T.-S. Kim, J. C. Hou, H. Lim, "Improving Spatial Reuse through Tuning Transmit Power, Carrier
Sense Threshold, and Data Rate in Multihop Wireless Networks,"
Procs., \emph{MOBICOM}, pp. 366-377, 2006.

\bibitem{etx} D. S. J. De Couto, D. Aguayo, J. Bicket, and R. Morris, "A high-throughput path metric for multi-hop wireless
routing," Procs., \emph{MOBICOM}, pp. 134-146, 2003.

\bibitem{wcett} R. Draves, J. Padhye, and B. Zill, "Routing in Multi-radio, Multi-hop Wireless Mesh Networks,"
Procs., \emph{MOBICOM}, 2004.

\bibitem{mtm} B. Awerbuch, D. Holmer, and H. Rubens, "The Medium Time Metric: High Throughput Route Selection
in Multi-rate Ad HocWireless Networks," \emph{Mobile Networks and
Applications}, vol. 11, pp. 253-266, 2006.

\bibitem{exor} S. Biswas, R. Morris, "ExOR: Opportunistic Multi-hop Routing for Wireless
Networks," Procs. \emph{SIGCOMM}, pp. 133-144, 2005.

\end{thebibliography}
\end{document}